\documentclass[aps,prd,twocolumn,nofootinbib,floatfix,eqsecnum]{revtex4-1} 

\usepackage{graphics}
\usepackage{graphicx}

\usepackage{latexsym}
\usepackage[cp866]{inputenc}

\input epsf

\begin{document}

\title{Strong isospin breaking at production of  light scalars
\footnote{The invited plenary talk presented by N.N. Achasov at the
14th International Workshop on Tau Lepton Physics, 19--25 September
2016, Beijing, China. To be published in Nuclear and Particle
Physics Proceedings.}}
\author{N.N. Achasov and G.N. Shestakov}
\affiliation{Laboratory of Theoretical Physics, S.L. Sobolev
Institute for Mathematics, 630090 Novosibirsk, Russia}

\begin{abstract}
It is discussed breaking the isotopic symmetry as the tool of
studying the production and nature of light scalar mesons.
\end{abstract}

\maketitle


\section{Introduction}
\label{Int}

The  thirty seven years ago we discovered theoretically a threshold
phenomenon known as the mixing of $a^0_0(980)$ and $f_0(980)$
resonances that breaks  the isotopic invariance considerably, since
the effect $\sim\sqrt{2(M_{K^0}-M_{K^+})/M_{K^0}}\approx 0,13$ in
the module of the amplitude \cite{ADS79}; see also Ref.
\cite{ADS81}. This effect appears as the narrow, $
2(M_{K^0}-M_{K^+})\approx 8$ MeV, resonant structure between the
$K^+K^-$ and $K^0\bar K^0$ thresholds, $a_0^0(980)\to K\bar K\to
f_0(980)$ and vice versa. Since that time many new proposals were
appeared, concerning both the searching it and estimating the
effects related with this phenomenon [3-29].

Nowadays, this phenomenon has been discovered experimentally and
studied with the help of detectors VES in Protvino \cite{Do08,Do11}
and BESIII in Beijing
\cite{Ab1,Ab2,Ab3} in the processes \\[0.2cm]
(a)\ \ \ $\pi^-N\to\pi^-f_1(1285)N\to\pi^-f_0(980)\pi^0N\to$

\vspace*{0.15cm} $\ \ \ \ \to$\,$\pi^-\pi^+\pi^-\pi^0N$ \ \cite{Do08,Do11}, \\[0.2cm]
(b)\ \ \ $J/\psi\to\phi f_0(980)\to\phi a_0(980)\to\phi\eta\pi^0$ \ \cite{Ab1}, \\[0.2cm]
(c)\ \ \ \,$\chi_{c1}$\,$\to$\,$a_0(980)\pi^0$\,$\to$\,$f_0(980)\pi^0$\,$\to$\,$\pi^+\pi^-\pi^0$ \ \cite{Ab1}, \\[0.2cm]
(d)\ \ \ $J/\psi\to\gamma\eta(1405)\to\gamma f_0(980)\pi^0\to\gamma\,3\pi$ \ \cite{Ab2}, \\[0.2cm]
(e)\ \ \ $J/\psi\to\phi f_0(980)\pi^0\to\phi\,3\pi$ \ \cite{Ab3}, \\[0.2cm]
(f)\ \ \ $J/\psi\to\phi f_1(1285)\to\phi f_0(980)\pi^0\to\phi\,3\pi$ \ \cite{Ab3} \\[0.2cm]
It has become clear \cite{AKS15,AKS16} that the similar isospin
breaking effect can appear not only due to the $a^0_0(980)-
f_0(980)$ mixing, but also for any mechanism of the production of
the $K\bar K$ pairs in the $S$ wave, $X\to K\bar K\to f_0(980)/
a^0_0(980)$.\,\footnote{Each such mechanism reproduces both the
narrow resonant peak and the sharp jump of the phase of the
amplitude between the $K^+K^-$ and $K^0\bar K^0$ thresholds.} Thus a
new tool to study the production mechanism and nature of light
scalars is emerged.

\section{\boldmath The $a^0_0(980)-f_0(980)$ mixing} \label{SecII}

The main contribution to the $a^0_0(980)-f_0(980)$ mixing amplitude,
caused by the diagrams shown in Fig. 1, has the form
\begin{figure}[!ht]
\hspace*{0.3cm}\includegraphics[width=17pc]{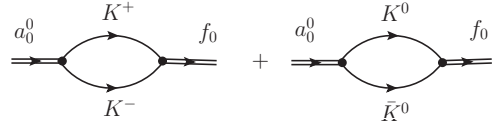}
\caption{\label{Figure1} The $K\bar K$ loop mechanism of the
$a^0_0(980)-f_0(980)$ mixing.}\end{figure}
\begin{eqnarray} \hspace*{-35pt}&&
\Pi_{a^0_0f_0}(m)=\frac{g_{a^0_0K^+K^-}g_{f_0K^+K^-
}}{16\pi}\Biggl[\,i\,\Bigl(\rho_{K^+K^-}(m)\nonumber\\
\hspace*{-35pt} &&
-\rho_{K^0\bar K^0}(m)\Bigr)- 
\frac{\rho_{K^+K^-}(m)}{\pi}\ln\frac{1+\rho_{K^+K^-}(m)}
{1-\rho_{K^+K^-}(m)}\nonumber\\ \hspace*{-35pt} && +\frac{\rho_{K^0
\bar K^0}(m)}{\pi}\ln\frac{1+\rho_{K^0\bar K^0}(m)}{1-\rho_{K^0\bar
K^0}(m)}\,\,\Biggl]\nonumber\\ \hspace*{-35pt} &&
\approx\frac{g_{a^0_0K^+K^-}g_{f_0K^+K^-
}}{16\pi}\,i\,\Bigl(\rho_{K^+K^-}(m)-\rho_{K^0\bar
K^0}(m)\Bigr),\nonumber\end{eqnarray} where $m$ (invariant virtual
mass of scalar resonances) $\geq2m_{K^0}$ and $\rho_{K\bar
K}(m)=\sqrt{1-4m_K^2/m^2}$; in the region $0\leq m\leq2m_K$,
$\rho_{K\bar K}(m)$ should be replaced by $i|\rho_{K\bar K}(m)|$.
The modulus and the phase of $\Pi_{a^0_0f_0}(m)$ are shown in Fig.
2.
\begin{figure}[!ht] \hspace{-0.35cm}
\includegraphics[width=20pc]{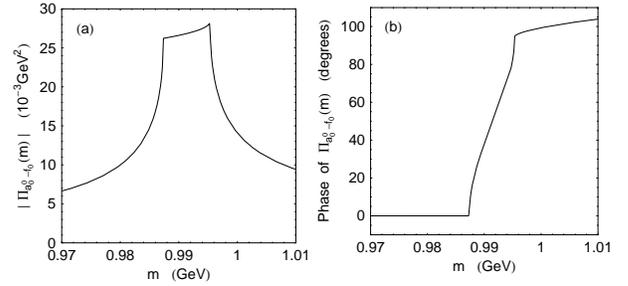}
\caption{\label{Figure2ab} (a) An example of the modulus of the
$a^0_0(980)-f_0(980)$ mixing amplitude. (b) The phase of the
$a^0_0(980)-f_0(980)$ mixing amplitude.}\end{figure}
In the region between the $K^+K^-$ and $K^0\bar K^0$ thresholds,
which is the 8\,MeV wide,
\begin{eqnarray}
\hspace*{-30pt} && |\Pi_{a^0_0f_0}(m)|\approx\frac{|g_{
a^0_0K^+K^-}g_{f_0K^+K^-}|}{16\pi}\sqrt{\frac{2(m_{K^0}-
m_{K^+})}{m_{K^0}}}\nonumber\\ \hspace*{-30pt} && \approx0.127
\frac{|g_{a_0K^+K^-}g_{f_0K^+K^-}|}{16\pi}\simeq0.03\mbox{ GeV}^2\nonumber\\
\hspace*{-30pt} &&\approx m_K\sqrt{m_{K^0}^2-m_{K^+}^2}\approx
m_K^{3/2}\sqrt{m_d-m_u}\,.\nonumber \end{eqnarray} Note that
$|\Pi_{\rho^0\omega}|\approx|\Pi_{\pi^0\eta}|\approx 0.003\mbox{
GeV}^2\approx (m_d-m_u)\times 1\mbox{ GeV}.$

The branching ratios of the isospin-breaking decays
$f_0(980)\to\eta\pi^0$ and $a^0_0(980)\to\pi^+\pi^-$, caused by the
$a^0_0(980)-f_0(980)$ mixing, are \cite{AKS16}
\begin{eqnarray}
BR(f_0(980)\to K\bar K\to a^0_0(980)\to\eta\pi^0)\nonumber\\
\hspace*{-8pt}=\int\left|\frac{\Pi_{a^0_0f_0}(m)}{D_{a^0_0}(m)
D_{f_0}(m)-\Pi^2_{a^0_0f_0}(m)}\right|^2\nonumber\\
\times\frac{2m^2\Gamma_{a^0_0 \to\eta\pi^0}(m)}{\pi}dm\approx
0.3\%\,,\ \ \ \nonumber \end{eqnarray}
\begin{eqnarray}
BR(a^0_0(980)\to K\bar K\to f_0(980)\to\pi\pi)\nonumber\\
\hspace*{-8pt}=\int\left|\frac{\Pi_{a^0_0f_0}(m)}{D_{a^0_0}(m)
D_{f_0}(m)-\Pi^2_{a^0_0f_0}(m)}\right|^2\nonumber\\
\times\frac{2m^2\Gamma_{f_0\to\pi\pi}(m)}{\pi}dm\approx 0.2\%\,,\ \
\ \nonumber\end{eqnarray} where $D_{a^0_0}(m)$ and $D_{f_0}(m)$ are
the propagators of the $a^0_0(980)$ and $f_0(980)$ resonances,
respectively.
\begin{figure}[!ht]
\hspace*{0.4cm}\includegraphics[width=15pc]{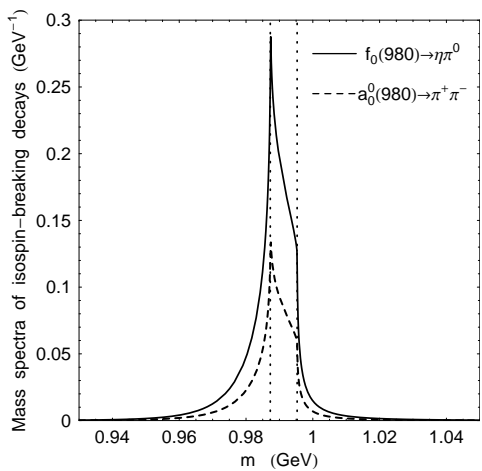}
\caption{\label{Fig1} Mass spectra in the isospin-violating decays
$f_0(980)\to\eta\pi^0$ and $a^0_0(980)\to\pi^+\pi^-$, caused by the
$a^0_0(980)-f_0(980)$ mixing. The solid and dashed lines are
generally similar each other. The dotted vertical lines show the
locations of the $K^+K^-$ and  $K^0\bar K^0$ thresholds.}
\end{figure}
Figure 3 shows the mass spectra correspond to the integrands in the
above equations.\,\footnote{Here we use the values of the coupling
constants of the $f_0(980)$ and $a^0_0(980)$ resonances with the
$\pi\pi$, $K\bar K$, and $\eta\pi$ channels obtained in Ref.
\cite{AKS16} from the BESIII data on the intensities of the
$f_0(980)\to a^0_0(980)$ and $a^0_0(980)\to f_0(980)$ transitions
measured in the reactions (b) and (c) \cite{Ab1}.}

\section{Polarization phenomena} \label{SecIII}

The phase jump (see Fig. 2(b)) suggests the idea to study the
$a_0^0(980)-f_0(980)$ mixing in polarization phenomena
\cite{AS04a,AS04b}. If the process amplitude with the spin
configuration is dominated by the $a_0^0(980)-f_0(980)$ mixing then
the spin asymmetry of the cross section jumps near the $K\bar K$
thresholds. An example is the reaction $\pi^-p_{\uparrow}\to\left
(a_0^0(980)+f_0(980)\right)n\to a_0^0(980)n\to\eta \pi^0\,n$ on a
polarized proton target. The corresponding differential cross
section has the form
\begin{eqnarray}
\frac{d^3\sigma}{dtdmd\psi}= \frac{1}{2\pi}\left[
\,|M_{++}|^2+|M_{+-}|^2\right.\nonumber\\ \left. + 2\,\Im
(M_{++}M^*_{+-})\,P \cos\psi\,\right],\nonumber\end{eqnarray} and
the dimensionless normalized spin asymmetry is $A(t,m)=2\,\Im
(M_{++}M^*_{+-})/(\,|M_{++}|^2 +|M_{+-}|^2\,)$, $\,-1\leq
A(t,m)\leq1$.\,\footnote{Here $M_{+-}$ and $M_{++}$ are the
$s$-channel helicity amplitude with ana without nucleon helicity
flip, $\psi$ is the angle between the normal to the reaction plain
formed by the momenta of the $\pi^-$ and $\eta\pi^0$ system, and the
transverse (to the $\pi^-$ beam axis) polarization of the the proton
target, and $P$ is a degree of this polarization.} Figure 4
illustrates the strong asymmetry jump which is straightforward
manifestation of the $a_0(980)-f_0(980)$ mixing amplitude
interfering with the isospin preserving one in the $\rho_2$ and
$\pi$ Regge exchange model. Details and various variants may be
found in Refs. \cite{AS04a,AS04b}.

These effects are still in waiting for their studies.
\begin{figure}
\hspace*{0.3cm}\includegraphics[width=15pc]{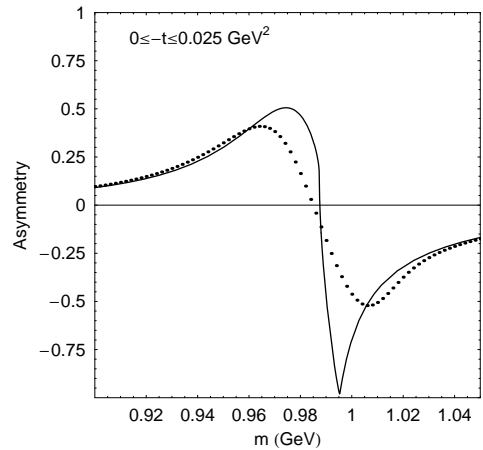}
\caption{\label{Fig2} Manifestation of the $a_0(980)-f_0(980)$
mixing effect in the reaction $\pi^-p_{\uparrow}\to a^0_0(980)
n\to\eta\pi^0n$ on a polarized target at $P^{\pi^-}_{lab}=18.3$ GeV
in the $\rho_2$ and $\pi$ exchange model. The solid (dotted) curves
show the spin asymmetry $A(0\leq t\leq0.025\mbox{ GeV}^2,m)$ as a
function of $\eta\pi^0$ invariant mass, $m$ (smoothed with 10 MeV
mass resolution).}\end{figure}

\section{\boldmath The decay $f_1(1285)\to f_0(980)\pi^0\to 3\pi$} \label{SecIV}

Estimated are the contributions of the following mechanisms
responsible for the  decay $f_1(1285)\to
f_0(980)\pi^0\to\pi^+\pi^-\pi^0$ \cite{AKS16}: \begin{description}
\item{(1)} the contribution of the $a^0_0(980)-f_0(980)$ mixing,
$f_1(1285)\to a_0(980)\pi^0\to (K^+K^-+K^0\bar K^0)\pi^0\to
f_0(980)\pi^0\to\pi^+\pi^-\pi^0$,\item{(2)} the contribution of the
transition  $f_1(1285)\to(K^+K^-+K^0\bar K^0)\pi^0\to f_0
(980)\pi^0\to\pi^+\pi^-\pi^0$, arising due to the pointlike decay
$f_1 (1285)\to K\bar K\pi^0$,\item{(3)} the contribution of the
transition $f_1(1285)\to(K^*\bar K+\bar K^*K)\to(K^+K^-+K^0\bar
K^0)\pi^0\to f_0(980)\pi^0\to\pi^+\pi^-\pi^0$, where $K^*=K^*(892)$,
and \item{(4)} the contribution of the transition
$f_1(1285)\to(K^*_0\bar K+\bar K^*_0K)\to(K^+K^-+K^0\bar
K^0)\pi^0\to f_0(980)\to\pi^+ \pi^-\pi^0$, where $K^*_0=K^*_0(800)$
(or $\kappa$) and $K^*_0(1430)$.\end{description} These mechanisms
break the conservation of the isospin  due to the nonzero mass
difference of the $K^+$ and $K^0$ mesons. They result in the
appearance of the narrow resonance structure in the $\pi^+\pi^-$
mass spectrum in the region of the $K\bar K$ thresholds, with the
width $\approx2m_{K^0}-2m_{K^+}\approx8$ MeV. The observation of
such a structure in experiment is the direct indication on the
$K\bar K$ loop mechanism of the breaking of the isotopic invariance.

We point out that existing data should be more precise, and it is
difficult to explain them using the single specific mechanism from
those listed above. Taking the decay $f_1(1285)\to f_0(980)\pi^0
\to\pi^+\pi^-\pi^0$ as the example, we discuss the general approach
to the description of the $K\bar K$ loop mechanism of the breaking
of isotopic invariance.

(1) The matter is that the $J/\psi$\,$\to$\,$\phi f_0(980)$
$\to$\,$\phi a_0(980)$\,$\to$\,$\phi\eta\pi^0$ \cite{Ab1} and
$\chi_{c1} $\,$\to$\,$a_0(980)\pi^0$\,$\to$ $ f_0(980)\pi^0$\,$
\to$\,$ \pi^+\pi^-\pi^0$ \cite{Ab1} decays are described by the
$a^0_0(980)-f_0(980)$ mixing well enough:
\begin{eqnarray}\label{EqIII-1}\hspace*{-35pt} &&
\frac{BR(J/\psi\to\phi f_0(980)\to\phi a^0_0(980)\to \phi\eta
\pi^0)}{BR(J/\psi\to\phi f_0(980)\to\phi\pi\pi)}\qquad \nonumber\\
\hspace*{-39pt} && =(0.60\pm0.20(stat.)\pm0.12(sys.) \pm0.26(para.)
)\%\nonumber\\ \hspace*{-35pt} && \approx\frac{BR(f_0(980)\to K\bar
K\to a^0_0(980) \to\eta\pi^0)}{BR(f_0(980)\to\pi\pi)}\,,\nonumber
\end{eqnarray}
\begin{eqnarray}\label{EqIII-2}\hspace*{-35pt} &&
\frac{BR(\chi_{c1}\to a^0_0(980)\pi^0\to f_0(980)\pi^0
\to\pi^+\pi^-\pi^0)}{BR(\chi_{c1}\to a^0_0(980)\pi^0\to\eta \pi^0
\pi^0)} \nonumber\\ \hspace*{-39pt} && =(0.31\pm0.16(stat.)
\pm0.14(sys.)\pm 0.03(para.))\%\nonumber\\ \hspace*{-35pt} &&
\approx\frac{BR(a^0_0(980)\to K\bar K\to f_0(980)\to\pi^+\pi^-)}
{BR(a^0_0(980)\to\eta\pi^0)}\,.\nonumber\end{eqnarray} As for the
$f_1(1285)\to f_0(980)\pi^0\to 3\pi$ decay \cite{Do11}, its
description requires the ``terrible''\ $a^0_0(980)-f_0(980)$ mixing:
\begin{eqnarray}\label{EqIII-3}\hspace*{-25pt} &&
\frac{BR(f_1(1285)\to a^0_0(980)\pi^0\to f_0(980)\pi^0
\to\pi^+\pi^-\pi^0)}{BR(f_1(1285)\to a^0_0(980)\pi^0\to\eta \pi^0
\pi^0)} \nonumber\\ \hspace*{-25pt} && =(2.5\pm0.9)\% \nonumber\\
\hspace*{-25pt} && \approx\frac{BR(a^0_0(980)\to K\bar K\to
f_0(980)\to\pi^+\pi^-)} {BR(a^0_0(980)\to\eta\pi^0)}\,,
\nonumber\end{eqnarray} and, as a result, the inconvenient coupling
constants of the scalar mesons with the pseudo-scalar mesons in the
many cases
\begin{eqnarray*}\frac{g^2_{f_0\pi^+\pi^-}}{4\pi}=1.2\mbox{\ GeV}^2,
\ \ \frac{g^2_{f_0 K^+K^-}}{4\pi}=5.7\mbox{\ GeV}^2,\\
\frac{g^2_{a^0_0\eta\pi^0}}{4\pi}=1.9\mbox{\ GeV}^2, \ \
\frac{g^2_{a^0_0 K^+K^-}}{4\pi}=9.9 \mbox{\ GeV}^2.\end{eqnarray*}
For example, due to the very strong coupling of $a^0_0(980)$ with
the $K\bar K$ channel, the width of the $a^0_0(980)$ resonance in
the $\eta\pi^0$ mass spectrum turns out to be near 15 MeV.

(2) The the pointlike decay $f_1(1285)\to K\bar K\pi^0$ gives
\begin{eqnarray*}\hspace*{-12pt}\frac{BR(f_1(1285)\to f_0(980)\pi^0
\to\pi^+\pi^-\pi^0)}{BR(f_1(1285)\to K\bar K \pi)}=0.0022
\end{eqnarray*}
instead of the experimental value
\begin{eqnarray*}\frac{BR(f_1(1285)\to f_0(980)\pi^0
\to\pi^+\pi^-\pi^0)}{BR(f_1(1285)\to K\bar K \pi)}\\
=0.033\pm 0.010\,.\qquad\qquad\qquad\end{eqnarray*} The $\pi^+\pi^-$
mass spectrum in the decay $f_1(1285)\to(K^+ K^-+K^0\bar
K^0)\pi^0\to f_0(980)\pi^0\to\pi^+ \pi^-\pi^0$ looks similar to the
curves in Fig. 3 for the $a_0(980)$-$f_0(980)$ mixing case. However,
it is clear that the pointlike mechanism of the decay $f_1(1285)\to
K\bar K\pi$ cannot by itself provide the considerable probability of
the $f_1(1285)\to f_0(980)\pi^0\to\pi^+\pi^-\pi^0$ transition.

(3) The isospin-breaking decay  $f_1(1285)\to(K^*\bar K+\bar
K^*K)\to( K^+K^-+K^0\bar K^0)\pi^0$\,$\to$\,$f_0(980)\pi^0$ $
\to$\,$\pi^+ \pi^-\pi^0$ is induced by the diagram shown in Fig. 5,
\begin{figure} [!ht] \hspace*{1.5mm}
\includegraphics[width=7.6cm]{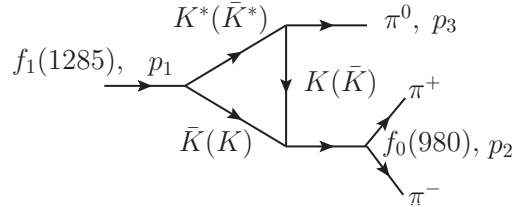}
\caption{\label{Fig3} The diagram of the decay $f_1(1285)\to
f_0(980)\pi^0\to\pi^+\pi^-\pi^0$ via the $K^*\bar K+\bar K^*K$
intermediate states.}\end{figure}
because the contributions from the $K^+K^-$ and  $K^0\bar K^0$ pair
production are not compensated entirely. The transition
$f_1(1285)\to(K^*\bar K+\bar K^*K)\to(K^+K^-+ K^0\bar K^0)\pi^0\to
f_0(980)\pi^0\to\pi^+\pi^-\pi^0$ gives the shape of the $\pi^+\pi^-$
spectrum practically coincides with the corresponding spectrum
caused by the $a^0_0(980)-f_0(980)$ mixing, but its
$$BR(f_1(1285)\to f_0(980)\pi^0\to\pi^+\pi^-\pi^0)\approx0.0255\%$$
is much less then the experimental value
\begin{eqnarray*}BR(f_1(1285)\to f_0(980)\pi^0
\to\pi^+\pi^-\pi^0)\\ =(0.30\pm 0.09)\%\,. \qquad
\qquad\qquad\end{eqnarray*} So, the $f_1(1285)\to(K^*\bar K+ \bar
K^*K)\to(K^+K^-+K^0\bar K^0)\pi^0\to f_0(980)\pi^0\to\pi^+
\pi^-\pi^0$ transition mechanism alone is also insufficient to
understand the experimental data.

(4) The variant $f_1(1285)\to(K^*_0(800)\bar K+ \bar K^*_0(800)K)\to
(K^+K^-+K^0\bar K^0)\pi^0\to f_0(980)\pi^0\to\pi^+ \pi^-\pi^0$ is
rejected by the shapes of the $K\pi$ and $K\bar K$ mass spectra in
the decay $f_1(1285)\to K\bar K\pi$. As for
$f_1(1285)\to(K^*_0(1430)\bar K+ \bar K^*_0(1430)K)\to (K^+K^-+
K^0\bar K^0)\pi^0\to f_0(980)\pi^0\to\pi^+\pi^-\pi^0$, it provides
the results similar to $f_1(1285)\to(K^*\bar K+ \bar K^*K)\to
(K^+K^-+K^0\bar K^0)\pi^0\to f_0(980)\pi^0\to\pi^+ \pi^-\pi^0$ and
consequently cannot describe the data alone.

\section{The consistency condition}

The isospin breaking amplitude $\mathcal{M}_{f_1(1285)\to
f_0(980)\pi^0}(m)$ can be expanded near the  $K\bar K$ threshold
into the series in $\rho_{K\bar K}(m)=\sqrt{1-4 m^2_K/m^2}$\,:
\begin{eqnarray}\label{EqVIII-2}
\hspace*{-25pt} & \mathcal{M}_{f_1(1285)\to
f_0(980)\pi^0}(m)=g_{f_0K^+K^-}\left\{A(m)\right.\ \qquad\nonumber\\
\hspace*{-25pt} & \times i[\rho_{K^+K^-}(m)-\rho_{K^0\bar K^0}(m)]
+B(m)[\rho^2_{K^+K^-}(m)\ \nonumber \\ \hspace*{-22.5pt} &
-\rho^2_{K^0 \bar K^0}(m)] +O[\rho^3_{K^+K^-}(m)-\rho^3_{K^0\bar
K^0}(m)]+\cdot\cdot\cdot \left.\right\}.\nonumber
\end{eqnarray} With a good accuracy
\begin{eqnarray*}
\mathcal{M}_{f_1(1285)\to f_0(980) \pi^0}(m)=g_{f_0K^+K^-}A(m)\\
\times i[\rho_{K^+K^-}(m)-\rho_{K^0\bar K^0}(m)].\ \ \qquad
\end{eqnarray*}
The amplitude $A(m)$ contains the information about all possible
mechanisms of production of the  $K\bar K$ system with isospin $I=1$
in $S$ wave in the process $f_1(1285)\to K\bar K\pi$.

From the data on the decay $f_1(1285)\to f_0(980)\pi^0\to\pi^+\pi^-
\pi^0$\, one can extract the information about  $|A(m)|^2$ in the
region of the $K^+K^-$ and $K^0\bar K^0$ thresholds,
\begin{eqnarray}\hspace*{10pt}\frac{d\Gamma_{f_1(1285)\to
f_0(980)\pi^0\to\pi^+ \pi^-\pi^0}(m)}{dm}\quad\nonumber \\
\hspace*{10pt}=\frac{1}{16\pi}|\mathcal{M}_{f_1(1285)\to
f_0(980)\pi^0}(m)|^2\quad\nonumber \\
\times p^3(m)\,\frac{2m^2\Gamma_{f_0\to\pi^+\pi^-}(m)}{
\pi|D_{f_0}(m)|^2}\,,\qquad \nonumber\end{eqnarray} where
$p(m)$\,=\,$[m^4_{f_1}-2m^2_{f_1}(m^2+m^2_\pi)+(m^2-m^2_\pi)^2]^{1/2}\\
/(2m_{f_1})$. Moreover, the information about $|A(m)|^2$ at $m>2m_K$
can be obtained from the data on the $K\bar K$ mass spectra measured
in the decays $f_1(1285)\to K\bar K\pi$. For instance, the $K^+K^-$
spectrum in the decay $f_1(1285)\to K^+K^-\pi^0$ can be represented
in the form
\begin{eqnarray}
\frac{d\Gamma_{f_1(1285)\to K^+K^-\pi^0}}{dm}\qquad\qquad \nonumber\\
\qquad =\frac{2\,m}{\pi}\,\rho_{K^+K^-}(m) \,p^3(m)\,|A(m)|^2\,.\ \
\nonumber\end{eqnarray} Fitting the data on $d\Gamma_{f_1\to
K^+K^-\pi^0}/dm$, one can find the value $|A (2m_{K^+})|^2$ and
obtain the following approximate estimate \cite{AKS16}
\begin{eqnarray}
\Gamma_{f_1(1285)\to f_0(980)\pi^0\to \pi^+\pi^-\pi^0}\ \ \ \ \nonumber\\
\qquad =|A (2m_{K^+})|^2\,2.59\times10^{-6}
\,\mbox{GeV}^5.\nonumber\end{eqnarray} Thus its comparison with the
data on the decay $f_1(1285)\to\pi^+\pi^-\pi^0$ permits one to
verify their consistence with the data on the decay  $f_1(1285)\to
K\bar K\pi$ and with the idea of the breaking of isotopic invariance
caused by the mass difference of  $K^+$ and  $K^0$ mesons.

\section{\boldmath The decay $J/\psi$\,$\to$\,$\gamma\eta(1405)$\,$\to$\,$\gamma
f_0(980)\pi^0\to\gamma\pi^+\pi^-\pi^0$}

According to BESIII \cite{Ab2}, the mass and the width of the
$\eta(1405)$ peak in the $\pi^+\pi^-\pi^0$ channel are
$1409.0\pm1.7$ MeV and $48.3\pm5.2$ MeV, respectively, while the
branching ratio is
\begin{eqnarray*}\hspace*{-16pt} BR(J/\psi\to\gamma\eta(1405)\to\gamma
f_0(980)\pi^0\to\gamma\pi^+\pi^-\pi^0)\nonumber
\\ =(1.50\pm0.11\pm0.11)\cdot10^{-5}\,.\qquad\quad\quad\
\end{eqnarray*}
In addition, the BESIII  gives the ratio \begin{eqnarray*}
\frac{BR(\eta (1405)\to f_0(980)\pi^0\to\pi^+\pi^-\pi^0)}{BR(\eta
(1405)\to a^0_0(980)\pi^0\to\eta\pi^0\pi^0)} \nonumber \nonumber
\\ =(17.9\pm4.2)\%\,,\qquad\qquad\qquad\end{eqnarray*}
that rules out practically the explanation of the discovered effect
by means of the $a_0(980)-f_0(980)$ mixing.
\begin{figure}[!ht]\hspace*{1.5mm}
\includegraphics[width=7.5cm]{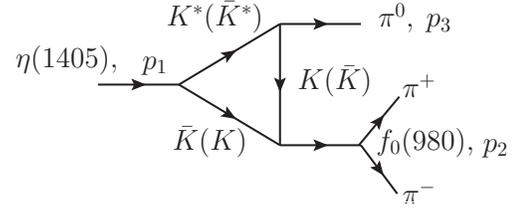}
\caption{\label{Fig5} The diagram of the decay $\eta(1405)\to(K^*
\bar K+\bar K^*K)\to K\bar K\pi\to\pi^+\pi^-\pi^0$. In the region of
the $\eta(1405)$ resonance all intermediate particles in the loop of
this triangle diagram can lie on their mass shells. That is, in the
hypothetical case of the stable $K^*$ meson the logarithmic
singularity appears in the imaginary part of the triangle
diagram.}\end{figure}

We discuss the possibility of the theoretical explanation of the
large breaking of isotopic invariance in the decay
$\eta(1405)$\,$\to$\,$f_0(980)\pi^0$\,$\to$\,$ \pi^+\pi^-\pi^0$ by
means of the anomalous Landau thresholds (the logarithmic triangle
singularities), which are in the transition $\eta(1405)\to(K^*\bar
K+\bar K^*K)\to(K^+K^-+K^0\bar K^0)\pi^0\to
f_0(980)\pi^0\to\pi^+\pi^-\pi^0$ (see Fig. 6), and show that the
account of the finite width of the $K^*(892)$ ($\Gamma_{K^*\to
K\pi}\approx50$ MeV) smoothes the logarithmic singularities in the
amplitude and results in the suppression of the calculated width of
the decay $\eta(1405)\to f_0(980)\pi^0\to3\pi$ by the factor of
$6-8$ in comparison with the case of $\Gamma_{K^*\to K\pi}=0$
\cite{AKS15}.

The accounting of the finite width of the $K^*$ resonance, i.e., the
averaging of the amplitude over the resonance Breit--Wigner
distribution in accord with the spectral K\"{a}ll\'{e}n--Lehmann
representation for the propagator of the unstable $K^*$ meson,
smoothes the logarithmic singularities of the amplitude and hence
makes the compensation of the contributions of the $K^{*+}K^-+
K^{*-}K^+$ and $K^{*0}\bar K^0+\bar K^{*0}K^0$ intermediate states
more strong. This results in both the diminishing of the calculated
width of the decay $\eta(1405)\to\pi^+\pi^-\pi^0$ by a number of
times in comparison with the case of $\Gamma_{K^*\to K\pi}=0$, and
in the concentration of the main effect of the isospin breaking in
the domain of the $\pi^+\pi^-$ invariant mass between the $K\bar K$
thresholds.

Assuming the dominance of the $\eta(1405)\to(K^*\bar K+\bar K^*K)\to
K\bar K\pi^0$ decay, one obtains
\begin{eqnarray*}\hspace*{-12pt}
BR(J/\psi\to\gamma\eta(1405)\to\gamma f_0(980)\pi^0\to\gamma 3\pi)
\\ \approx1.12\cdot10^{-5}\,,\qquad\qquad\qquad\end{eqnarray*}
that reasonably agrees with experiment.

$${\small\mbox{ VI\,a.}}\ \ {\it Conclusion}$$

We also analyze the difficulties related with the assumption of the
dominance of the $\eta(1405)\to(K^*\bar K+\bar K^*K)\to K\bar K\pi$
decay mechanism and discuss the possible dynamics of the decay
$\eta(1405)\to\eta\pi\pi$ \cite{AKS15}. The decisive improvement of
the experimental data on the $K\bar K$, $ K\pi$, $\eta\pi$, and
$\pi\pi$ mass spectra in the decay of the resonance structure
$\eta(1405/1475)$ to $K\bar K\pi$ and $\eta\pi\pi$, and on the shape
of the resonance peaks themselves in the $K\bar K\pi$ and
$\eta\pi\pi $ decay channels is necessary for the further
establishing the $\eta(1405)\to3\pi$ decay mechanism.

\begin{center}{\small\bf ACKNOWLEDGMENTS}\end{center} 

The present work is partially supported by the Russian Foundation
for Basic Research Grant No. 16-02-00065 and the Presidium of the
Russian Academy of Sciences project No. 0314-2015-0011.

\end{document}